\documentclass[%
 preprint,
nofootinbib,
 amsmath,amssymb,
 prc,
 superscriptaddress
]{revtex4-1}

\usepackage{graphicx}
\usepackage{dcolumn}
\usepackage{bm}
\usepackage{caption}
\usepackage{subcaption}
\usepackage{color}

\begin{document}


\title{$\alpha$ Decays in Superstrong Static Electric Fields}


\author{Dong Bai}
\email{dbai@itp.ac.cn}
\affiliation{School of Physics, Nanjing University, Nanjing, 210093, China}%

\author{Zhongzhou Ren}
\email{zren@tongji.edu.cn}
\affiliation{School of Physics Science and Engineering, Tongji University, Shanghai 200092, China}%


\date{\today}

\begin{abstract}
Superstrong static electric fields could deform Coulomb barriers between $\alpha$ clusters and daughter nuclei, and bring up the possibility of speeding up $\alpha$ decays. We adopt a simplified model for the spherical $\alpha$ emitter $^{212}$Po and study its responses to superstrong static electric fields. We find that, superstrong electric fields with field strengths $|\mathbf{E}|\sim0.1$ MV/fm could turn the angular distribution of $\alpha$ emissions from isotropic to strongly anisotropic, and speed up $\alpha$ decays by more than one order of magnitude. We also study the influences of superstrong electric fields along the Po isotope chains, and discuss the implications of our studies on $\alpha$ decays in superstrong monochromatic laser fields. The study here might be helpful for future theoretical studies of $\alpha$ decay in realistic superstrong laser fields.
\end{abstract}

\maketitle



$\alpha$ decay occupies an important position in nuclear physics. Its explanation in terms of quantum tunneling by Gamow, Condon, and Gurney was one of the first applications of quantum mechanics in nuclear physics \cite{Gamow:1928,Gurney:1928}. After that, many attempts have been made to give accurate descriptions of $\alpha$ decays from both phenomenological and microscopic viewpoints. See, e.g., Refs.~\cite{Delion:2010,Delion:2018rrl} and references therein.
New strong laser sources in the future may create new opportunities for $\alpha$-decay studies \cite{ELI,ELINP}. One of the most important open questions is whether strong lasers can be used to speed up $\alpha$ decays, which, if feasible, would bring up new possibility for decontaminating $\alpha$-radioactive transuranium nuclear wastes \cite{Cortes:2011,Cortes:2012tu,Misicu:2013,Kopytin:2014,Misicu:2016,Delion:2017ozx,Kis:2018llv,Bai:2018adq}. In this work, instead of handling directly monochromatic or more realistic tightly-focused laser fields, we pursuit first the possibility of changing $\alpha$-decay half-lives by using superstrong static electric fields. Static electric fields correspond to the low-frequency limit of monochromatic laser fields in the dipole approximation (see, e.g., Ref.~\cite{Mittleman:1993,Joachain:2011}). Thanks to its intrinsic theoretical simplicity, the study of $\alpha$ decays in static electric fields can be used as a benchmark for understanding $\alpha$ decays in the presence of monochromatic laser fields with finite frequencies and tightly focused laser fields, providing us with valuable opportunities to establish correct physical pictures and gain useful physical intuitions.    


In this article, we adopt the simplified model proposed in Ref.~\cite{Delion:2009jw} to capture the main features of $\alpha$ decays from even-even $\alpha$ emitters. The $\alpha$-core potential is given by
\begin{align}
V(r)&=\frac{1}{2}\mu_\alpha\Omega^2(r-r_0)^2+V_0,\qquad r\le r_B,\nonumber\\
&=\frac{Z_cZ_de^2}{r},\qquad r>r_B,
\label{DelionModel}
\end{align}
corresponding to join the shifted harmonic oscillator potential to the Coulomb potential at $r=r_B$. $r$ is the $\alpha$-core relative distance. $\mu_\alpha=M_cM_d/(M_c+M_d)$ is the two-body reduced mass. $Z_c$ and $Z_d$ are the charges of the $\alpha$ cluster and the core nucleus, respectively. $r_0=1.2A_d^{1/3}$ is the core radius. $r_B$ is taken to be the touching radius $r_B=1.2(A_{c}^{1/3}+A_{d}^{1/3})$. When $r_0\gg\sqrt{\frac{\hbar}{\mu_\alpha\Omega}}$, the initial quasi-stable (QS) state (the emission energy $Q_\alpha$) of $\alpha$ decays corresponds approximately to the first eigenstate (the first eigenvalue $\frac{1}{2}\hbar\Omega$) of the shifted harmonic oscillator well. With the help of the continuity condition of $V(r)$ at the top of the potential barrier $r_B$, we have
\begin{align}
&Q_\alpha-V_0=\frac{1}{2}\hbar\Omega,\\
&\frac{1}{2}\mu_\alpha\Omega^2(r_0-r_B)^2+V_0=\frac{Z_cZ_de^2}{r_B},
\end{align}
from which we obtain the following relation
\begin{align}
\frac{1}{2}\mu_\alpha\Omega^2(r_0-r_B)^2=V_\text{frag}(r_B)+\frac{1}{2}\hbar\Omega.
\end{align}
$V_\text{frag}(r)={Z_cZ_de^2}/{r}-Q_\alpha$ is the so-called fragmentation potential.
The simplified model given by Eq.~\eqref{DelionModel} simulates the Pauli principle by the hard core provided by the left wing of the shifted harmonic oscillator well, and allows us to reproduce the important fact that the preformation amplitude of the $\alpha$ cluster has an approximate Gaussian profile centered on the surface of the core nucleus \cite{Varga:1992zz,Delion:1992zz,Ropke:2014wsa,Xu:2015pvv}. 
Moreover, it is shown in Ref.~\cite{Delion:2009jw} that the simplified model of Eq.~\eqref{DelionModel} could give a simple explanation of the Viola-Seaborg rule for $\alpha$-decay half-lives \cite{Viola:1966}.

When turning on external superstrong static electric fields, the Hamiltonian in the center-of-mass (CM) frame is given by
\begin{align}
\left[-\frac{\hbar^2}{2\mu_\alpha}\nabla_r^2+V(r)-eZ_\text{eff}\mathbf{E}\cdot\mathbf{r}\right]\Psi(\mathbf{r})=\widetilde{Q}_\alpha\Psi(\mathbf{r}),
\end{align}
which is analogous to the Stark Hamiltonian in atomic physics. $\widetilde{Q}_\alpha$ is the new emission energy for $\alpha$ decays in superstrong static electric fields and can be estimated within, e.g., the framework of the two-potential method \cite{Gurvitz:1987,Gurvitz:1988}. The interactions between nuclei and electric fields are depicted in the so-called length gauge. The electric field $\mathbf{E}$ is treated as classical. In this article, we consider electric fields that are strong enough to cause sizable effects in $\alpha$ decays, but not too strong to even diminish Coulomb barriers and thus drastically change the picture of quantum tunneling, or cause relativistic effects, i.e., we require $|\mathbf{E}|\ll \text{Min}\left\{E_\text{ob}\approx\frac{Z_cZ_de^2-Q_\alpha r_B}{eZ_\text{eff}r_B^2},E_\text{rel}\approx\frac{Q_\alpha\mu_\alpha c^2}{e^3 Z_\text{eff}Z_cZ_d}
\right\}$, with $E_\text{ob}$ and $E_\text{rel}$ being the over-barrier and relativistic thresholds. $Z_\text{eff}=(Z_cM_d-Z_dM_c)/(M_c+M_d)$ is the effective charge of the $\alpha$ cluster in the CM frame \cite{Misicu:2013}. The appearance of the effective charge $Z_\text{eff}$ could be understood intuitively as follows. In the laboratory frame, the accelerations of the $\alpha$ cluster and the core nucleus induced by the electric field are given by $\mathbf{a}_{c,\text{L}}=eZ_c\mathbf{E}/M_c$ and $\mathbf{a}_{d,\text{L}}=eZ_d\mathbf{E}/M_d$, respectively. When transformed into the CM frame,  the acceleration of the $\alpha$ cluster is given instead by ${\mathbf{a}}_{c,\text{CM}}=eZ_\text{eff}\mathbf{E}/M_c$, which could be interpreted as that the $\alpha$ cluster has an effective charge $Z_\text{eff}$ rather than the bare charge $Z_c$ in the CM frame. In nuclear physics, the differences between effective and bare charges of $\alpha$ clusters are, in general, not tiny. This should be contradicted with atomic physics, where, thanks to the huge mass hierarchy between electrons and nuclei ($M_c\ll M_d$), the differences between the effective and bare charges of electrons turn out to be so tiny that they could often be ignored safely.

The $\alpha$-decay width is generally parametrized by \cite{Delion:2010}
\begin{align}
\Gamma_\alpha=S_\alpha\gamma_\alpha^2\mathcal{P}_\alpha,
\end{align}
with $S_\alpha$ being the spectroscopic factor, $\gamma^2_\alpha$ being the reduced decay width, and $\mathcal{P}_\alpha$ being the barrier penetrability. Superstrong static electric fields, when presented, could impact $\alpha$ decays in the following ways at least:
\begin{enumerate}
\item Superstrong static electric fields could deform the potential barrier $V(r)$ through the additional interaction term $-eZ_\text{eff}\mathbf{E}\cdot\mathbf{r}$. Consequently, the spectroscopic factor $S_\alpha$, the reduced width $\gamma^2_\alpha$, and the barrier penetrability $\mathcal{P}_\alpha$ could all be changed. 
%
%
\item The presence of superstrong static electric fields introduces a preferred direction, i.e., the direction parallel to the electric field strength $\mathbf{E}$, into the problem. Therefore, anisotropic effects take place.
\item Superstrong static electric fields also change the subsequent motion of the emitted $\alpha$ particle, accelerating it continuously if the superstrong static electric fields extend infinitely in the space. This byproduct effect on emitted $\alpha$ particles could also be generalized to monochromatic and tightly focused laser fields. For example, monochromatic laser fields drive the emitted $\alpha$ particle back and forth, and may make it even recollide against the core nucleus \cite{Cortes:2012tu}. 
\end{enumerate}
Of course, superstrong static electric fields might also generate copious numbers of electron-positron pairs due to the QED effects in the vaccuum, as well as induce other changes to the inner structures of target nuclei. The impacts of these effects on $\alpha$ decays have been hardly investigated in literature yet and lie beyond the scope of the current work. Therefore, we ignore them for simplicity in the following.  

In this article, we shall concentrate on influences of static electric fields on the barrier penetrability $\mathcal{P}_\alpha$, and leave influences on the spectroscopic factor $S_\alpha$ and the reduced width $\gamma^2_\alpha$ to future works. It is well-known from the quantum mechanical perturbation theory that, in the Stark effect the first-order correction (proportional to $eZ_\text{eff}|\mathbf{E}|$) to $Q_\alpha$ vanishes if the target system has an inversion symmetry \cite{Friedrich:2017}. By dimensional analysis, the second-order correction to $Q_\alpha$ is estimated to be $\Delta Q_\alpha^{(2)}\sim\frac{(eZ_\text{eff}|\mathbf{E}|r_0)^2}{\hbar\Omega}$. As a result, we have approximately $\widetilde{Q}_\alpha\approx Q_\alpha$ for alpha-decay systems if $\Delta Q_\alpha^{(2)}\ll Q_\alpha$, i.e., $|\mathbf{E}|\ll E_{\Delta Q}=\frac{\sqrt{Q_\alpha\hbar\Omega}}{eZ_\text{eff}r_0}$. The total barrier penetrability $\widetilde{\mathcal{P}}_\alpha$ can be expressed in the semi-classical approximation by
\begin{align}
&\widetilde{\mathcal{P}}_\alpha=\frac{1}{2}\int_0^\pi\widetilde{\mathcal{P}}_\alpha(\Theta)\sin\Theta\,\mathrm{d}\Theta,\\
&\widetilde{\mathcal{P}}_\alpha(\Theta)=\exp\left(-2\int_{r_2(\Theta)}^{r_3(\Theta)}\sqrt{\frac{2\mu_\alpha}{\hbar^2}[V(r)-eZ_\text{eff}|\mathbf{E}|r\cos\Theta-Q_\alpha]}\,\,\mathrm{d}r\right),\label{DifferentialPenetrability}
\end{align}
with $\widetilde{\mathcal{P}}_\alpha(\Theta)$ as the differential barrier penetrability, and $r_2(\Theta)$ and $r_3(\Theta)$ as the internal and external turning points,
\begin{align}
&\!\!\!r_2(\Theta)=\frac{r_0\mu_\alpha\Omega^2+e |\mathbf{E}| Z_\text{eff} \cos\Theta+\sqrt{\hbar\mu_\alpha\Omega^3+2eZ_\text{eff}|\mathbf{E}|r_0\mu_\alpha\Omega^2\cos\Theta+e^2Z_\text{eff}^2|\mathbf{E}|^2\cos\Theta^2}}{\mu_\alpha\Omega^2},\\
&\!\!\!r_3(\Theta)=\frac{-Q_\alpha+\sqrt{Q_\alpha^2+4e^3Z_cZ_dZ_\text{eff}|\mathbf{E}|\cos\Theta}}{2eZ_\text{eff}|\mathbf{E}|\cos\Theta}.
\end{align}
It is important to note that, for $|\mathbf{E}|>E_\Theta=\frac{Q_\alpha^2}{4e^3 Z_c Z_d Z_\text{eff}}$, the angle $\Theta$ can only take values between $0$ and $\Theta_E=\arccos\left(-\frac{Q_\alpha^2}{4e^3Z_cZ_dZ_\text{eff}|\mathbf{E}|}\right)$. In other words, superstrong static electric fields may forbid $\alpha$ emissions in directions with $\Theta>\Theta_E$. This is a drastic manifestation of the anisotropic effects induced by superstrong static electric fields. 

\begin{figure}[tb]
\centering
\includegraphics[width=0.8\textwidth]{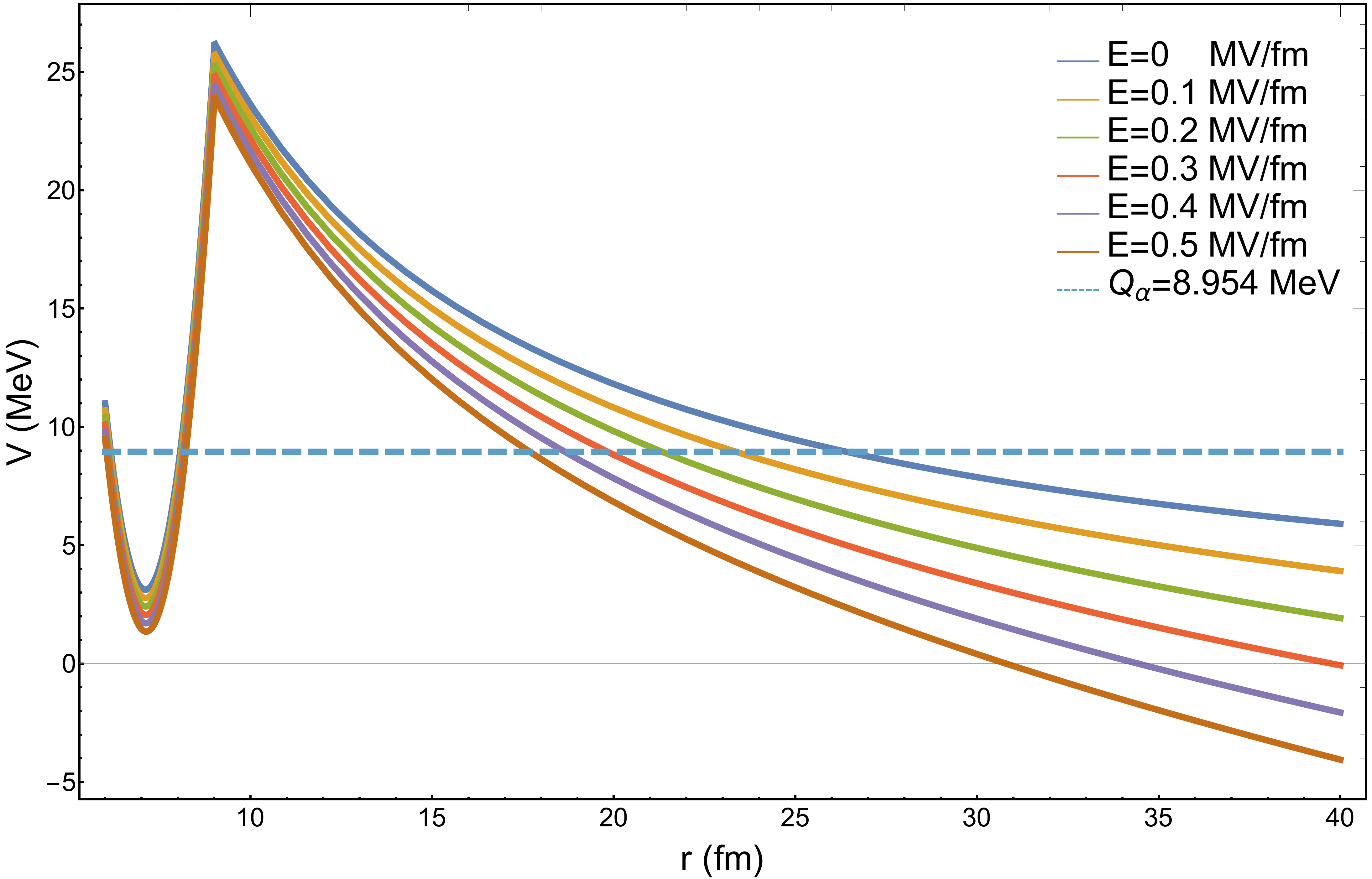}
\caption{The potential energy $V(r)$ between the $\alpha$ cluster and the core nucleus $^{208}$Pb along the direction parallel to the electric field $\mathbf{E}$ $(\Theta=0)$ with different electric fields ranging from $E=0$ MV/fm to $E=0.5$ MV/fm. The dashed line denotes the emission energy $Q_\alpha=8.954$ MeV.}
\label{Potential} 
\end{figure}

\begin{figure}[tb]
\centering
\includegraphics[width=0.8\textwidth]{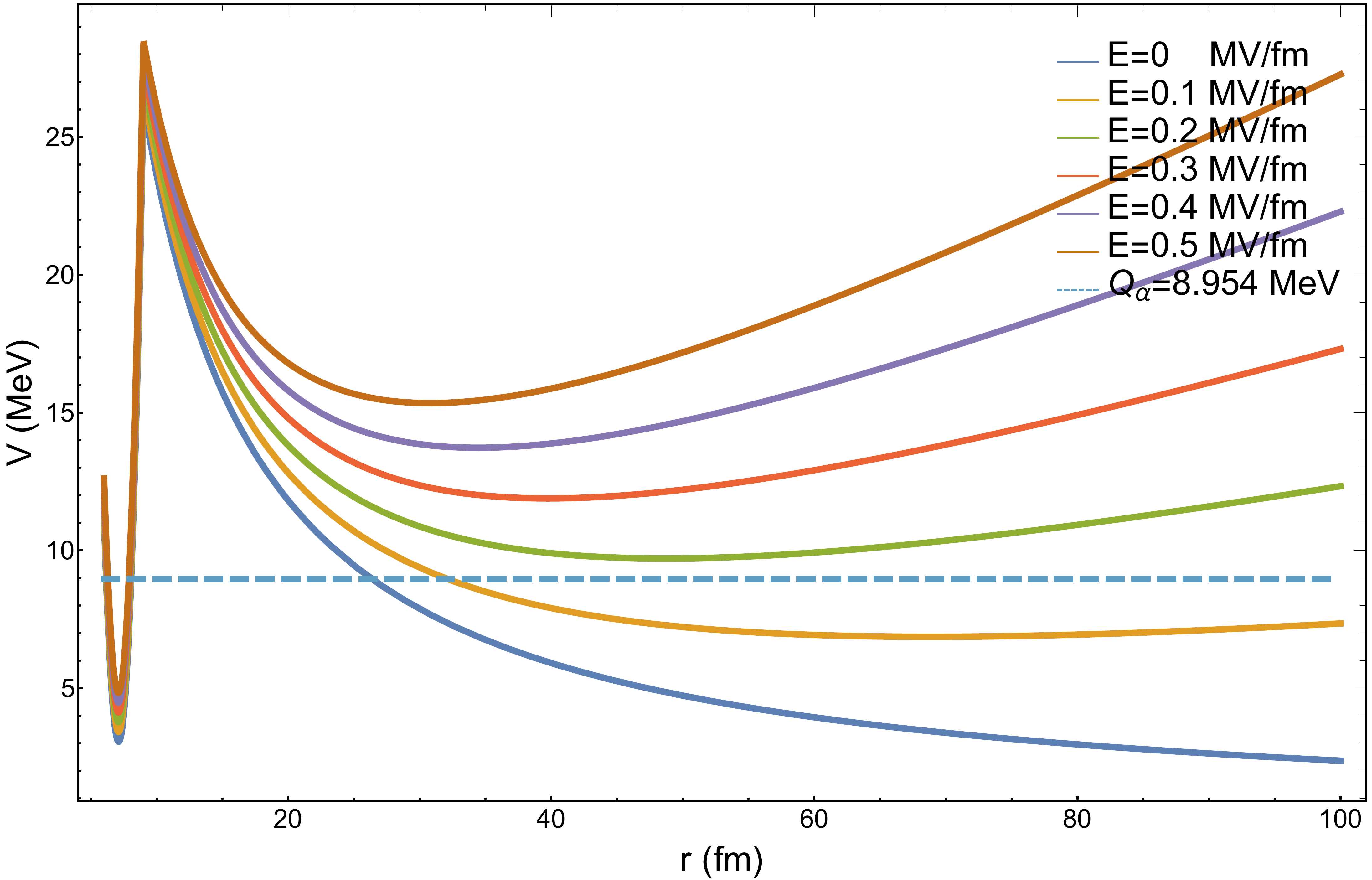}
\caption{The same as Fig.~\ref{Potential}, except that the $\alpha$-emission direction is anti-parallel to the electric field, i.e., $\Theta=\pi$. }
\label{PotentialOpposite} 
\end{figure}

Let's take the spherical $\alpha$ emitter $^{212}$Po $(^{208}$Pb$+\alpha)$ as an example to see explicitly the effects of superstrong static electric fields. The effective charge is given by $Z_\text{eff}=0.4151$, clearly different from the bare charge of $\alpha$ cluster $Z_c=2$. Various aforementioned characteristic field strengths are given by $E_\text{ob}=3.84$ MV/fm, $E_\text{rel}=278.34$ MV/fm, $E_{\Delta Q}=2.88$ MV/fm, and $E_{\Theta}=0.17$ MV/fm for the $^{212}$Po system. In this article, we shall mainly concentrate on superstrong static electric fields ranging from 0 to 0.5 MV/fm, satisfying the requirements $|\mathbf{E}|\ll\text{Min}\left\{E_\text{ob},E_\text{rel},E_{\Delta Q}\right\}$. Therefore, the $\alpha$-decay process could still be described by the quantum-tunneling picture of non-relativistic quantum mechanics with the emission energy being approximately unchanged. $\alpha$ decays with electric fields stronger than the threshold strengths are generally much more complicated, and lie beyond our present scope. The potentials $V(r)$ modified by superstrong static electric fields with different field strengths are plotted in Fig.~\ref{Potential} and \ref{PotentialOpposite}, with the $\alpha$-emission direction being parallel and anti-parallel to the electric field. Also, we plot a horizontal dashed line representing $Q_\alpha=8.954$ MeV, the emission energy of $^{212}$Po. It is important to article that, as shown in Fig.~\ref{PotentialOpposite}, the external turning point $r_3(\Theta=\pi)$ disappears for $|\mathbf{E}|>E_{\Theta}=0.17$ MV/fm.

\begin{figure}[tb]
\centering
\includegraphics[width=0.8\textwidth]{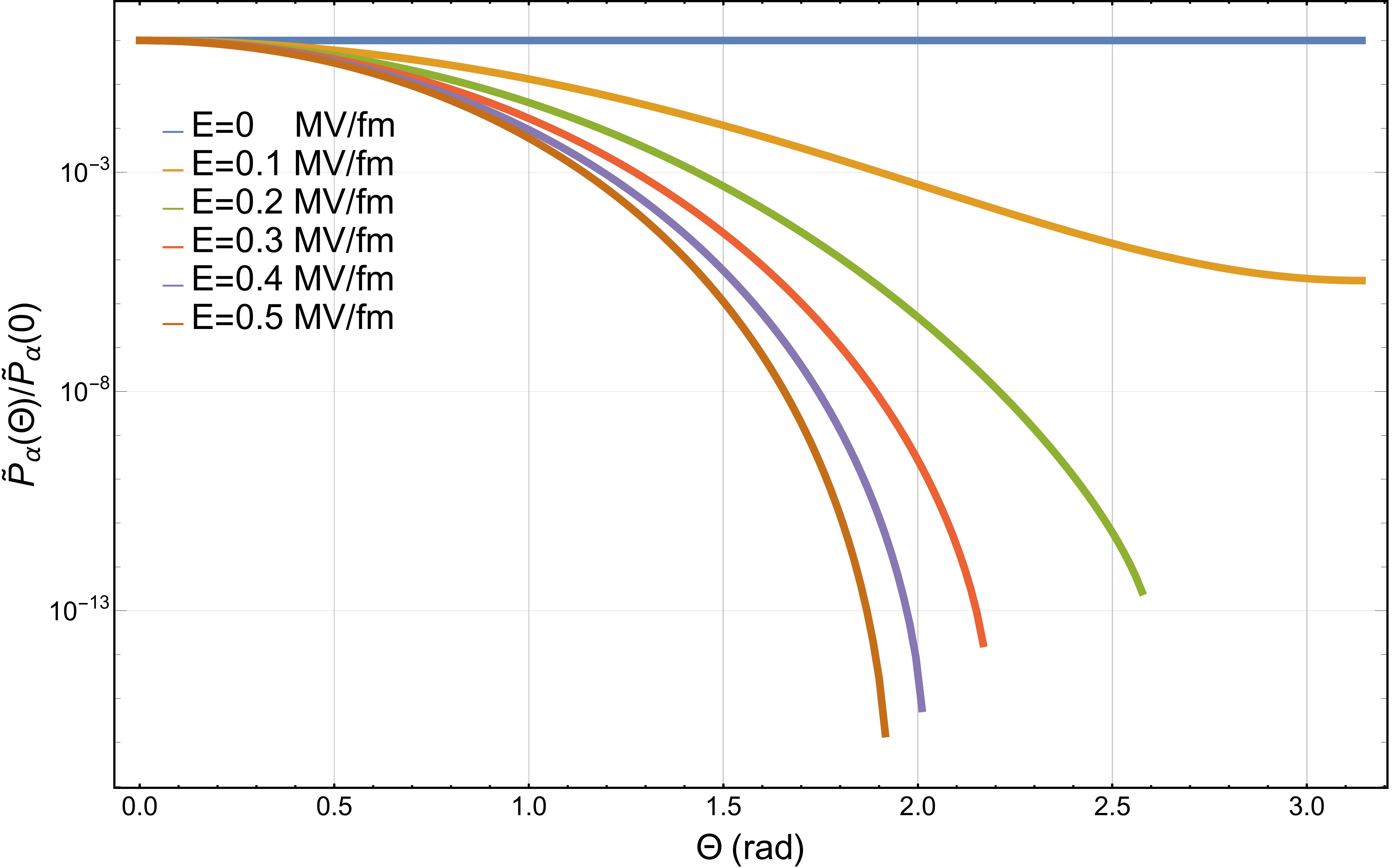}
\caption{Differential Penetrability for $\alpha$ decays in superstrong static electric fields with field strengths ranging from 0 to 0.5 MV/fm.}
\label{DifferentialPenetrabilityFig} 
\end{figure}

In Fig.~\ref{DifferentialPenetrabilityFig}, we plot the differential barrier penetrability $\widetilde{P}_\alpha(\Theta)$ normalized by its maximal value at $\Theta=0$ versus the angle $\Theta$ between the $\alpha$-emission direction $\mathbf{r}$ and the static electric field $\mathbf{E}$. We have several observations. First, angular distributions of $\alpha$ decays from spherical $\alpha$ emitters become anisotropic in superstrong static electric fields. The anisotropy increases strongly from a factor of about six orders of magnitude at $E=0.1$ MV/fm to more than sixteen orders of magnitude at $E=0.5$ MV/fm, with $\alpha$ emissions practically focused in the direction parallel to the electric field, i.e., $\mathbf{r}\parallel\mathbf{E}$. Second, for superstrong static electric field strengths with $|\mathbf{E}|>E_{\Theta}=0.17$ MV/fm, not all the $\Theta$ values are available for $\alpha$ emissions. This is revealed in Fig.~\ref{DifferentialPenetrabilityFig} by the termination of the $\tilde{P}_\alpha(\Theta)$ curves before $\Theta$ reaches $\pi$. Explicitly, we have $\Theta_E=3.14,3.14,2.59,2.17,2.01,1.92$ rad for $|\mathbf{E}|=0,0.1, 0.2, 0.3, 0.4, 0.5$ MV/fm, respectively. It might be sensible to imagine that $\Theta_E\to\pi/2$ when $|\mathbf{E}|$ becomes sufficiently large, as inspired by the definition of $\Theta_E$.

\begin{figure}[tb]
\centering
\includegraphics[width=0.8\textwidth]{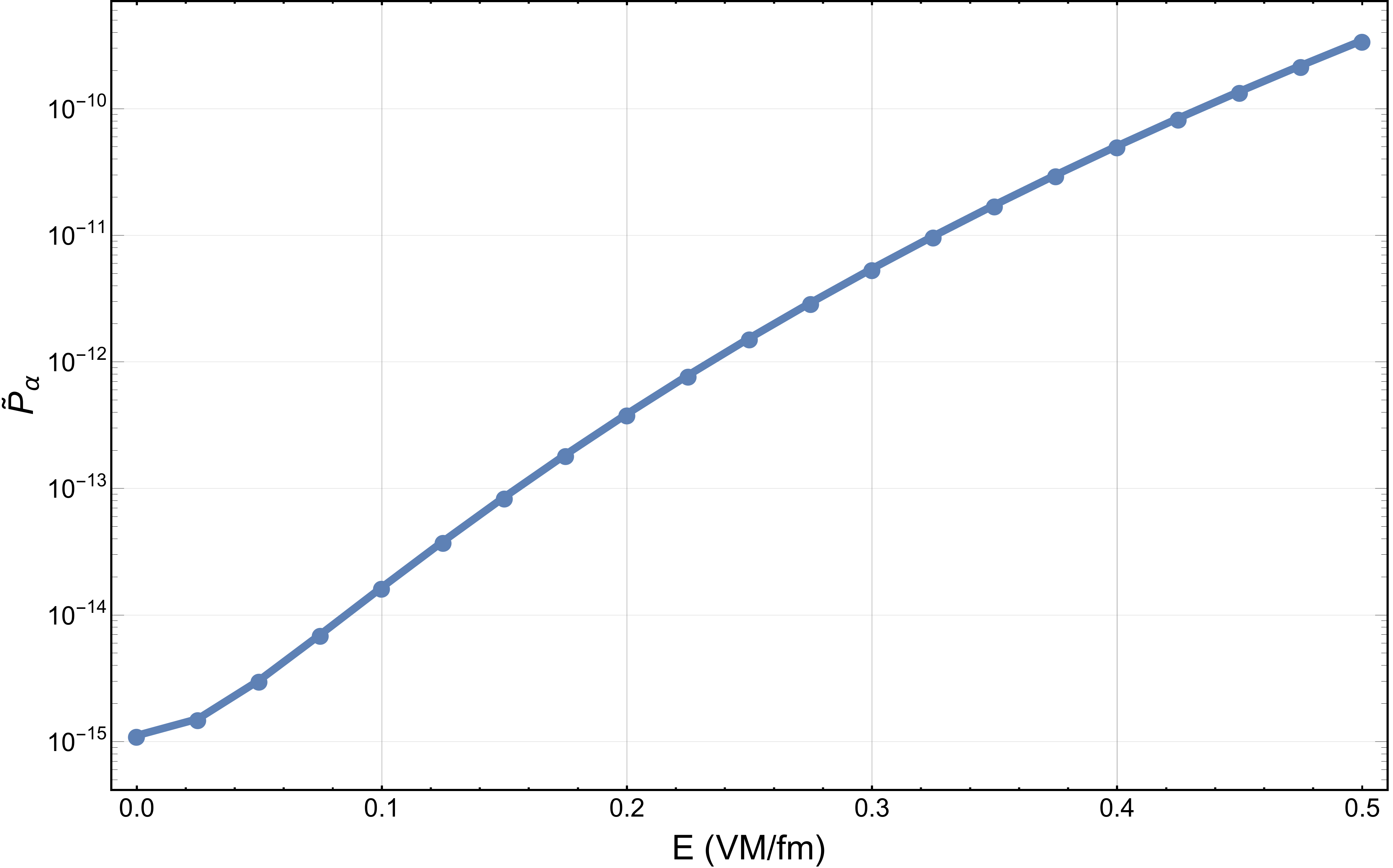}
\caption{Total penetrability $\widetilde{P}_\alpha$ versus the electric field strength $|\mathbf{E}|$ for the spherical $\alpha$ emitter $^{212}$Po. The electric field strength $|\mathbf{E}|$ goes from 0 to 0.5 MV/fm.}
\label{TotalPenetrabilityStrongerElectricFields} 
\end{figure}

\begin{figure}[tb]
\centering
\includegraphics[width=0.8\textwidth]{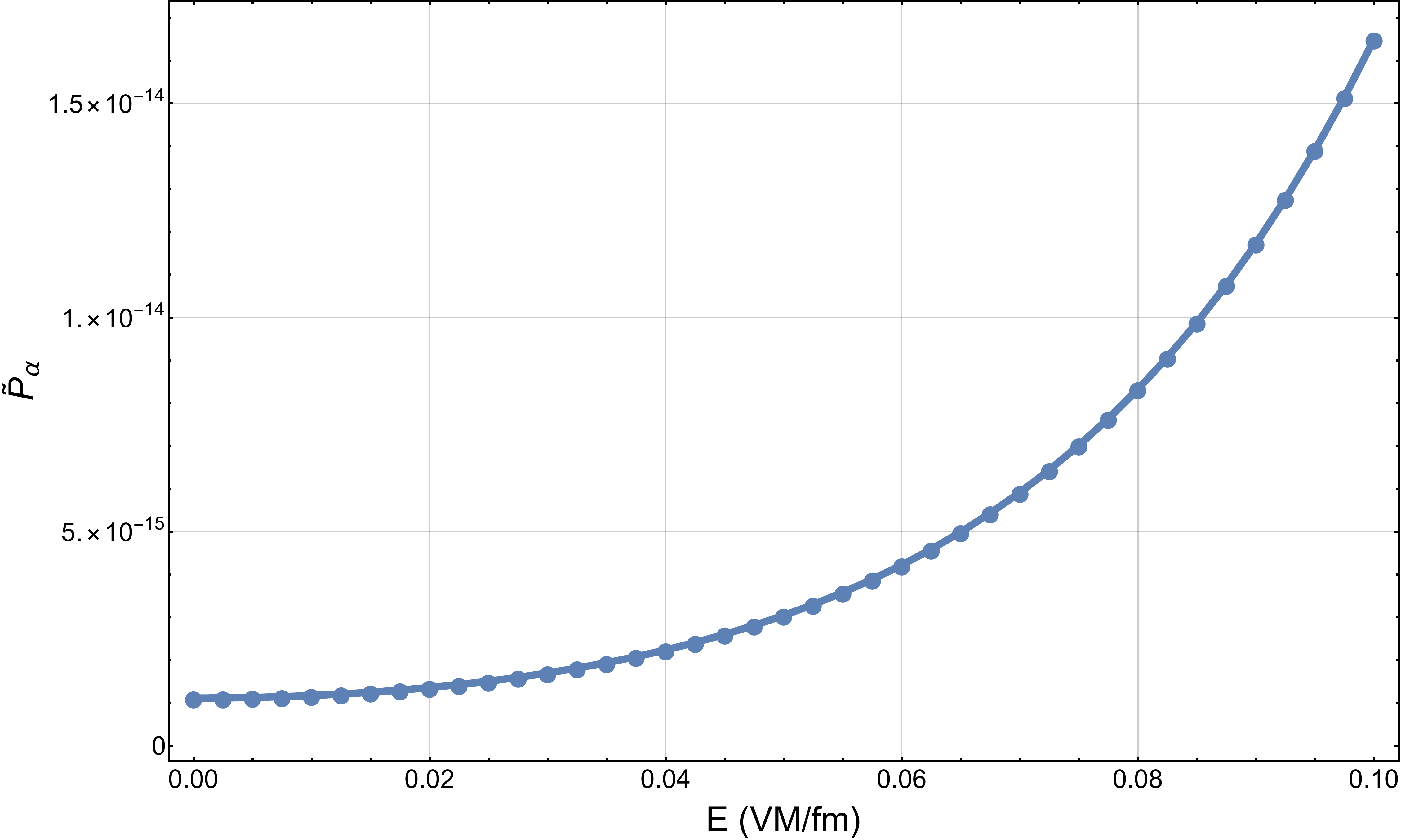}
\caption{The magnification of Fig.~\ref{TotalPenetrabilityStrongerElectricFields} for the electric field strength $|\mathbf{E}|$ ranging from 0 to 0.1 MV/fm. The total penetrability $\widetilde{P}_\alpha$ starts to increase wildly only when $|\mathbf{E}|$ exceeds the critical value of $E_\text{cr}\sim0.05$ MV/fm.}
\label{TotalPenetrabilityWeakerElectricFields} 
\end{figure}

We then study how the total penetrability $\widetilde{P}_\alpha$ changes with the electric field strength $|\mathbf{E}|$. In Fig.~\ref{TotalPenetrabilityStrongerElectricFields}, the electric field strength $|\mathbf{E}|$ is chosen to range from 0 to 0.5 MV/fm. It is straightforward to see that superstrong static electric fields could enhance greatly the total penetrability $\widetilde{P}_\alpha$. For $|\mathbf{E}|\sim 0.1$ MV/fm, the enhancement is over one order of magnitude, while for $|\mathbf{E}|\sim 0.5$ MV/fm, the enhancement is about five orders of magnitude. Suppose the spectroscopic factor $\widetilde{S}_\alpha$ and the reduced width $\widetilde{\gamma}_\alpha^2$ remain to be roughly of the same order of magnitude. Our calculations show clearly that superstrong static electric fields could speed up $\alpha$ decays, thus might be helpful for decontaminating transuranium nuclear wastes with long-living $\alpha$ emitters. Fig.~\ref{TotalPenetrabilityWeakerElectricFields} magnifies the behavior of $\widetilde{P}_\alpha$ versus $|\mathbf{E}|\sim 0-0.1$ MV/fm, from which we could see further that $\widetilde{P}_\alpha$ starts to increase wildly only when $|\mathbf{E}|$ exceeds the critical value $E_\text{cr}\sim0.05$ MV/fm.  

\begin{figure}[tb]
\centering
\includegraphics[width=0.9\textwidth]{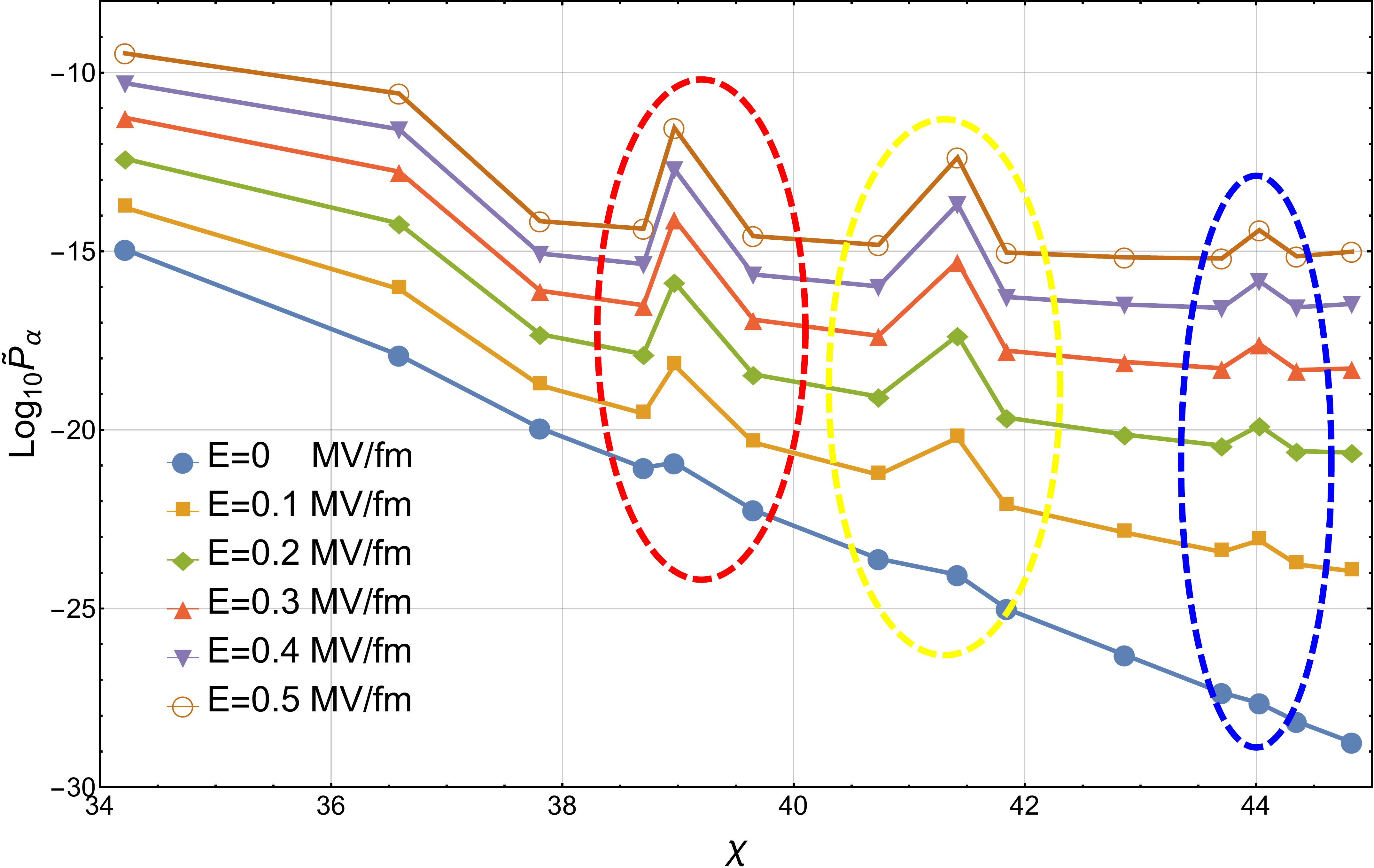}
\caption{Logarithm of the total penetrability $\widetilde{P}_\alpha$ vs the Coulomb-Sommerfeld parameter $\chi$ for the Po isotope chain. We consider different values of the electric field strength $|\mathbf{E}|$, ranging from 0 to 0.5 MV/fm. The celebrating GN law is reproduced for the $|\mathbf{E}|=0$ case, with several small bumps along the straight line. As $|\mathbf{E}|$ increases, these bumps start to inflate (enclosed in the dashed ellipses). }
\label{GeigerNuttallLaw} 
\end{figure}

We analyze further the total penetrability along the isotope chain of even-even Po isotopes with $A=192-218$. For simplicity, all nuclei are treated as spherical. The numerical results are given by Fig.~\ref{GeigerNuttallLaw}. The horizontal axis denotes the value of the Coulomb-Sommerfeld parameter $\chi=\frac{2Z_cZ_de^2}{\hbar\sqrt{\frac{2Q_\alpha}{\mu_\alpha}}}$, while the vertical axis denotes the logarithm of the total penetrability $\text{Log}_{10}\widetilde{P}_\alpha$. The electric field strength $|\mathbf{E}|$ ranges from 0 to 0.5 MV/fm. The $|\mathbf{E}|=0$ case reproduces explicitly the celebrating Geiger-Nuttall (GN) law for the $\alpha$ decay \cite{Geiger:1911}, which says that
\begin{align}
\text{Log}_{10}T_{1/2}=a\chi+b(Z).
\end{align}
In Fig.~\ref{GeigerNuttallLaw}, colorful dashed ellipses enclose small bumps corresponding to deviations from the ideal GN law. These bumps inflate as $|\mathbf{E}|$ increases, because the $\alpha$ emitter at the top has the larger effective charge than its neighbors. Take $\alpha$ emitters enclosed in the red ellipse as an example. For left to right, we have $^{194}$Po, $^{216}$Po, and $^{196}$Po. The $Q$ values of these three $\alpha$ emitters are close to each other, being 6.987 MeV,  6.906 MeV, and 6.658 MeV, while effective charges are given by 0.268, 0.444, and 0.286, respectively. It is then straightforward to see that $^{216}$Po with the larger effective charge could give the larger response to the external electric field, resulting in the growth of the bump in Fig.~\ref{GeigerNuttallLaw}. Others bumps in Fig.~\ref{GeigerNuttallLaw} could be explained in a similar way.

\begin{figure}[tb]
\centering
\includegraphics[width=0.8\textwidth]{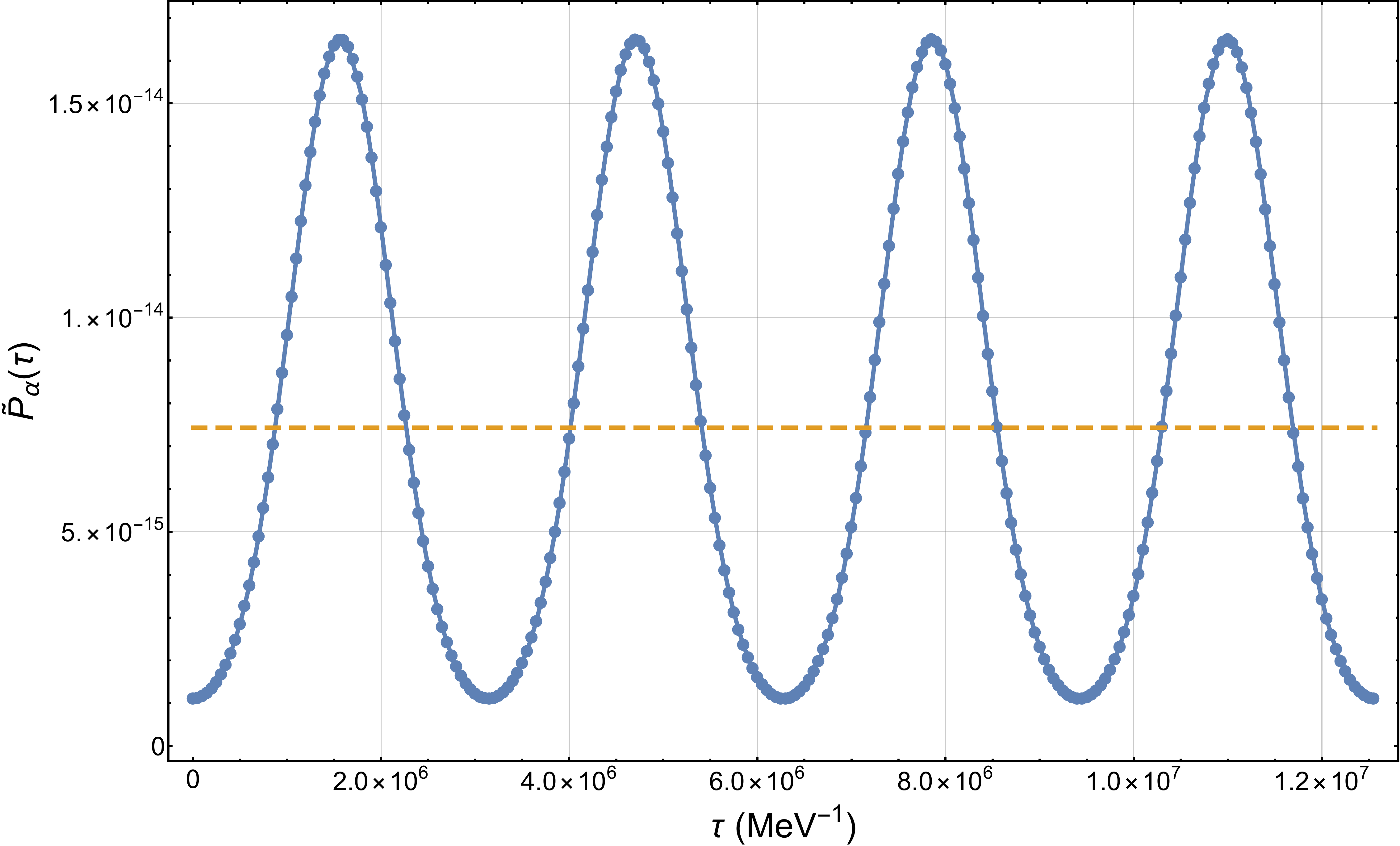}
\caption{Time-dependent barrier penetrability $\widetilde{P}_\alpha(\tau)$ versus the reduced time $\tau=t/\hbar$ for the alpha emitter $^{212}$Po. The electric field amplitude is taken to be $\mathcal{E}=0.1$ MV/fm. The photon energy is taken to be $\hbar\omega=1$ eV. The average barrier penetrability is displayed by the dashed horizontal line and turns to be $\widetilde{P}_{\alpha,\text{av}}=7.433\times10^{-15}$.}
\label{BarrierPenetrabilityAV} 
\end{figure}

\begin{figure}[tb]
\centering
\includegraphics[width=0.8\textwidth]{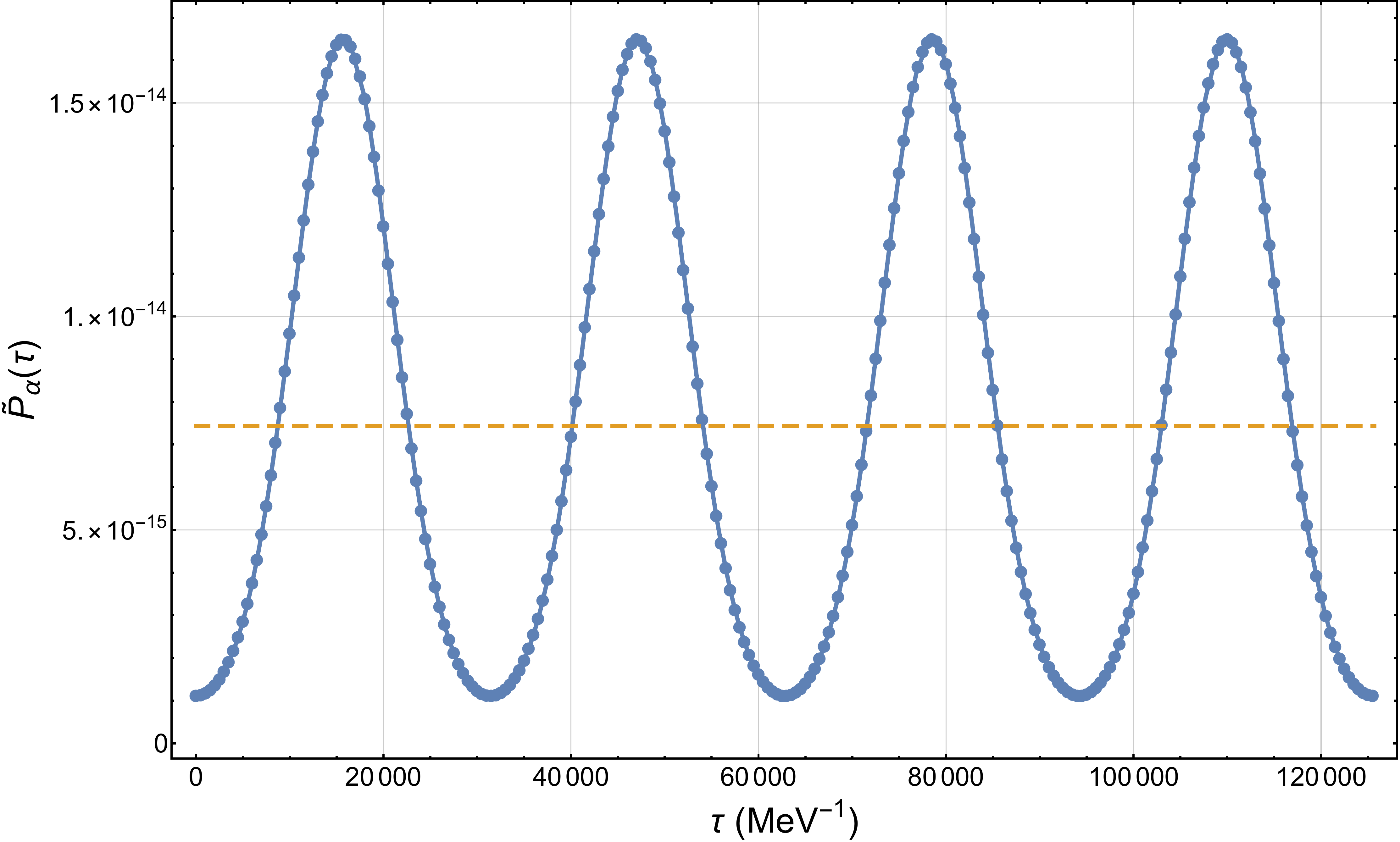}
\caption{The same as Fig.~\ref{BarrierPenetrabilityAV}, except that the photon energy is taken to be $\hbar\omega=100$ eV. The average barrier penetrability is then found to be $\widetilde{P}_{\alpha,\text{av}}=7.433\times10^{-15}$ as well.}
\label{BarrierPenetrabilityAVXRay} 
\end{figure}

Last but not least, we would like to comment briefly on the implications of our present study on $\alpha$ decays in monochromatic laser fields. In this article, we mainly discuss $\alpha$ decays in superstrong static electric fields. Similar to laser-atom physics, our  results could also be applied to monochromatic laser fields in the dipole approximation, which are closer to realistic laser fields generated by modern laser facilities, as long as the so-called adiabatic condition is obeyed \cite{Mittleman:1993,Joachain:2011}. We introduce $U_p=\frac{e^2Z_c^2\mathcal{E}^2}{4\mu_\alpha\omega^2}$, which is the pondermotive energy of an $\alpha$ particle in a monochromatic laser field with the laser frequency being $\omega$ and the electric-field amplitude being $\mathcal{E}$. The adiabatic condition could then be formulated as
\begin{align}
\gamma_\text{K}=\sqrt{\frac{V_\text{frag}(r_B)}{U_p}}\ll1\Longrightarrow\hbar\omega\ll\frac{\hbar eZ_c\mathcal{E}}{\sqrt{4\mu_\alpha V_\text{frag}(r_B)}},
\end{align}   
with $V_\text{frag}(r_B)$ being the ``ionization energy'' for $\alpha$ decays. Roughly speaking, this condition expresses the requirement that the laser field remains essentially static while the $\alpha$ cluster undergoes the tunneling process. In other words, the $\alpha$ decay is treated as adiabatic if the time taken by the $\alpha$ particle to tunnel is shorter than the time during which the laser field changes significantly. In laser-atom physics, $\gamma_\text{K}$ is also known as the Keldysh parameter \cite{Keldysh:1965}. For $^{212}$Po, the adiabatic condition gives that 
$\hbar\omega\ll0.943\,\mathcal{E}\text{ MeV}$,
which could be satisfied easily by most optical and X-ray laser sources with the electric field amplitude $\mathcal{E}\sim 0.1$ MV/fm. Therefore, for $\alpha$ decays in superstrong optical and X-ray laser fields, the average barrier penetrability could be obtained approximately by \cite{Mittleman:1993,Joachain:2011}
\begin{align}
{\widetilde{P}}_{\alpha,\text{av}}=\frac{1}{T}\int_0^{T}\widetilde{P}_\alpha[\mathbf{E}(t)]\mathrm{d}t=\frac{\hbar\omega}{2\pi}\int_0^{\frac{2\pi}{\hbar\omega}}\widetilde{P}_\alpha[\mathcal{E}\sin(\hbar\omega\tau)]\mathrm{d}\tau.
\label{AlphaDecayMonochromaticLaser}
\end{align}
Here, we take the laser electric field to be $\mathbf{E}(t)=\mathcal{E}\sin(\omega t)=\mathcal{E}\sin(\hbar\omega\tau)$. The reduced time $\tau=t/\hbar$ is introduced for convenience. Fig.~\ref{BarrierPenetrabilityAV} plots the time-dependent barrier penetrability $\widetilde{P}_\alpha(\tau)$ versus the reduced time $\tau$ for the $\alpha$ emitter $^{212}$Po. The electric field amplitude is taken to be $\mathcal{E}=0.1$ MV/fm, and the photon energy of the laser field is taken to be $\hbar\omega=1$ eV, corresponding to optical laser fields. Then, according to Eq.~\eqref{AlphaDecayMonochromaticLaser}, the average barrier penetrability turns out to be $\widetilde{P}_{\alpha,\text{av}}=7.433\times10^{-15}$. Fig.~\ref{BarrierPenetrabilityAVXRay} is almost the same as Fig.~\ref{BarrierPenetrabilityAV}, except that the laser photon energy is chosen to be $\hbar\omega=100$ eV, corresponding to X-ray laser fields. The average barrier penetrability remains the same in both cases, revealing its weak dependence on the laser frequency in the low-frequency region. 

In this article, we study $\alpha$ decays in superstrong static electric fields. The spherical $\alpha$ emitter $^{212}$Po is taken as an explicit example. It is found that superstrong electric fields with field strength as large as 0.1 MV/fm could deform significantly the potential between the $\alpha$ cluster and the core nucleus, making the angular distribution of the $\alpha$ emission become strongly anisotropic and speeding up $\alpha$ decays. The critical electric field strength that marks the wild growth of $\alpha$-decay widths is found to be $E_\text{cr}\sim0.05$ MV/fm for $^{212}$Po. We then study the effects of superstrong electric fields on the even-even Po isotope chain, and find that the bumps corresponding to deviations from the idealized GN law start to inflate when the electric field strength increases gradually. Last but not least, we also study the implications of our results on $\alpha$ decays in superstrong monochromatic lasers. In this work, we concentrate mainly on the impacts of superstrong electric fields on the barrier penetrability. It is also important to study the behaviors of the spectroscopic factor and reduced width in superstrong electric fields in future works, which, if available, would allow a more accurate understanding of the present topic. Producing superstrong laser fields with the electric field strength as large as, say., $0.05-0.1$ MV/fm is, of course, not an easy task \cite{Fedotov:2010,Nerush:2011,Piazza:2012,Narozhny:2015,Tamburini:2017}. One possible technical routes might be based on the idea of laser-nucleus collision \cite{Burvenich:2006zp}, where the target nucleus is accelerated to the ultra-relativistic speed, and collides against strong laser beams. Then, special relativity demands that, the electric field strength in the nucleus rest frame could be magnified correspondingly compared with that in the laboratory frame. For giant accelerators like the Large Hadron Collider (LHC) in Europe \cite{LHC}, or the proposed Super Proton-Proton collider (SppC) in China \cite{CEPC}, this magnification factor could be as large as about $15000-100000$, and thus enhance the electric field strength from, e.g., 1 V/fm in the lab frame, which is the planned electric field strength of Extreme Light Infrastructure - Nuclear Physics (ELI-NP) in Europe \cite{ELINP}, to 0.015$-$0.1 MV/fm in the nucleus rest frame. According to our studies, electric fields with such strength might be able to modify $\alpha$ decays significantly. The study here might be helpful for future theoretical studies on $\alpha$ decay in realistic superstrong laser fields.

\begin{acknowledgments} 
This work is supported by the National Natural Science Foundation of China (Grant No.~11535004, 11761161001, 11375086, 11120101005, 11175085 and 11235001), by the National Key R\&D Program of China (Contract No.~2018YFA0404403), by the National Major State Basic Research and Development of China, Grant No.~2016YFE0129300, and by the Science and Technology Development Fund of Macau under Grant No.~008/2017/AFJ.
\end{acknowledgments}


\begin{thebibliography}{999}

\bibitem{Gamow:1928}
G.~Gamow, 
 Z.\ Phys.\ {\bf 51}, 204 (1928).
 
 \bibitem{Gurney:1928}
 R.~Gurney, and E.~Condon, 
 Nature (London) {\bf 122}, 439 (1928).
 
 \bibitem{Delion:2010}
 D.~S.~Delion, 
 \emph{Theory of Particle and Cluster Emission} 
 (Springer-Verlag, Berlin, 2010).
 
\bibitem{Delion:2018rrl} 
  D.~S.~Delion, Z.~Ren, A.~Dumitrescu and D.~Ni,
  J.\ Phys.\ G {\bf 45}, 053001 (2018).
  
%
%
%
%
%
%
%
%
%
%
%
%
%
%
%
%
%
%
%
%
%
%
%
%
%
%
%
%
%
%
  
  
    \bibitem{ELI}
  Extreme Light Infrastructure (ELI), www.eli-laser.eu.
  
  \bibitem{ELINP}
  Extreme Light Infrastructure - Nuclear Physics (ELI-NP), www.eli-np.ro.
  
  \bibitem{Cortes:2011}
  H.~M.~Casta$\tilde{\text{n}}$eda Cortes,
  Doctoral thesis, University of Heidelberg (2011).
  
\bibitem{Cortes:2012tu} 
  H.~M.~Casta$\tilde{\text{n}}$eda Cortes, C.~Muller, C.~H.~Keitel and A.~Palffy,
  Phys.\ Lett.\ B {\bf 723}, 401 (2013)
  [arXiv:1207.2395 [nucl-th]].
      
  \bibitem{Misicu:2013}
  S.~Misicu and M.~Rizea, 
  J.\ Phys.\ G {\bf 40}, 095101 (2013).
  
      \bibitem{Kopytin:2014}
  I.~V.~Kopytin and A.~S.~Kornev, 
  Phys.\ At.\ Nucl.\ {\bf 77}, 53 (2014).
  
  \bibitem{Misicu:2016}
  S.~Misicu and M.~Rizea, 
  Open Phys.\ {\bf 14}, 81 (2016).
  
\bibitem{Delion:2017ozx} 
  D.~S.~Delion and S.~A.~Ghinescu,
  Phys.\ Rev.\ Lett.\  {\bf 119}, 202501 (2017).
  
\bibitem{Kis:2018llv} 
  D.~P.~Kis and R.~Szilvasi,
  J.\ Phys.\ G {\bf 45}, 045103 (2018).
  
\bibitem{Bai:2018adq} 
  D.~Bai, D.~Deng and Z.~Ren,
  Nucl.\ Phys.\ A {\bf 976}, 23 (2018)
  [arXiv:1805.02379 [nucl-th]].
  
\bibitem{Delion:2009jw} 
  D.~S.~Delion,
  Phys.\ Rev.\ C {\bf 80}, 024310 (2009)
  [arXiv:0907.2304 [nucl-th]].

\bibitem{Varga:1992zz} 
  K.~Varga, R.~G.~Lovas and R.~J.~Liotta,
  Phys.\ Rev.\ Lett.\  {\bf 69}, 37 (1992).
  
\bibitem{Delion:1992zz} 
  D.~S.~Delion, A.~Insolia and R.~J.~Liotta,
  Phys.\ Rev.\ C {\bf 46}, 1346 (1992).

\bibitem{Ropke:2014wsa} 
  G.~R\"opke {\it et al.},
  Phys.\ Rev.\ C {\bf 90}, 034304 (2014)
  [arXiv:1407.0510 [nucl-th]].
  
\bibitem{Xu:2015pvv} 
  C.~Xu {\it et al.},
  Phys.\ Rev.\ C {\bf 93}, 011306 (2016)
  [arXiv:1511.07584 [nucl-th]].
 
  \bibitem{Mittleman:1993}
  M.~H.~Mittleman, 
  \emph{Introduction to the Theory of Laser-atom Interactions}, 
  second edition (Plenum Press, New York, 1993).
  
      \bibitem{Joachain:2011}
  C.~J.~Joachain, N.~J.~Kylstra, and R.~M.~Potvliege, 
  \emph{Atoms in intense laser fields} (Cambridge University Press, Cambridge, 2011).
  
  \bibitem{Viola:1966}
  V.~E.~Viola, G.~T.~Seaborg, 
  J.~Inorg.~Nucl.~Chem.~{\bf 28}, 741 (1966).
  
  \bibitem{Gurvitz:1987}
 S.~A.~Gurvitz and G.~Kalbermann, 
 Phys.\ Rev.\ Lett.\ {\bf 59}, 262 (1987).
 
 \bibitem{Gurvitz:1988}
 S.~A.~Gurvitz,
 Phys.\ Rev.\ A {\bf 38}, 1747 (1988).
 
 \bibitem{Friedrich:2017}
 H.~Friedrich, 
 \emph{Theoretical Atomic Physics},
 fourth edition (Springer International Publishing AG, 2017).

\bibitem{Geiger:1911}
H.~Geiger and J.~M.~Nuttall, 
Philos.\ Mag.\ {\bf 22}, 613 (1911).

\bibitem{Keldysh:1965}
L.~V.~Keldysh,
Sov.\ Phys.\ JETP {\bf 20}, 1307 (1965).

\bibitem{Fedotov:2010}
A.~Fedotov, N.~Narozhny, G.~Mourou, and G.~Korn, 
Phys.\ Rev.\ Lett.\ {\bf 105}, 080402 (2010).

\bibitem{Nerush:2011}
E.~Nerush, I.~Kostyukov, A.~Fedotov, N.~Narozhny, N.~Elkina, and H.~Ruhl,
Phys.\ Rev.\ Lett.\ {\bf 106}, 035001 (2011).

\bibitem{Piazza:2012}
A.~Di Piazza, C.~M\"uller, K.~Z.~Hatsagortsyan, and C.~H.~Keitel, 
 Rev.\ Mod.\ Phys.\ {\bf 84}, 1177 (2012).
 
 \bibitem{Narozhny:2015}
 N.~B.~Narozhny, and A.~M.~Fedotov,  
  Contemp.\ Phys.\ {\bf 56}, 249 (2015).
  
  \bibitem{Tamburini:2017}
M.~Tamburini, A.~Di Piazza, C.~H.~Keitel,
Sci.\ Rep.\ {\bf 7}, 5694 (2017).

\bibitem{Burvenich:2006zp} 
  T.~J.~Burvenich, J.~Evers and C.~H.~Keitel,
  Phys.\ Rev.\ Lett.\  {\bf 96}, 142501 (2006)
  [nucl-th/0601077].
  
  \bibitem{LHC}
Large Hadron Collider, homepage at https://home.cern/topics/large-hadron-collider.

\bibitem{CEPC}
Circular Electron-Positron Collider, homepage at http://cepc.ihep.ac.cn.






\end{thebibliography}
\end{document}